\documentclass[aps,prb,twocolumn,showpacs,amsmath,amssymb,superscriptaddress,floatfix]{revtex4-1}
\usepackage{graphicx}
\usepackage{dcolumn}
\usepackage{bm}
\usepackage{amsfonts}
\usepackage{amsmath}
\usepackage{color}
\usepackage{hyperref}
\usepackage[babel=true]{csquotes}

\DeclareMathOperator{\arccosh}{arccosh}

\begin{document}

\title{Topological Josephson ${\phi}_0$-junctions}

\author{Fabrizio Dolcini}
\affiliation{Dipartimento di Scienza Applicata e Tecnologia del Politecnico di Torino, I-10129 Torino, Italy}
\affiliation{CNR-SPIN, Monte S.Angelo - via Cinthia, I-80126 Napoli, Italy}
\email{fabrizio.dolcini@polito.it} 
\author{Manuel Houzet}
\affiliation{Univ.~Grenoble Alpes, INAC-SPSMS, F-38000 Grenoble, France;\\ CEA, INAC-SPSMS, F-38000 Grenoble, France}
\author{Julia S. Meyer}
\affiliation{Univ.~Grenoble Alpes, INAC-SPSMS, F-38000 Grenoble, France;\\ CEA, INAC-SPSMS, F-38000 Grenoble, France}

\begin{abstract}
We study the effect of a magnetic field on the current-phase relation of a topological Josephson junction formed by connecting two superconductors through the helical edge states of a quantum spin-Hall insulator. We predict that the Zeeman effect along the spin quantization axis of the helical edges results in an anomalous Josephson relation that allows for a supercurrent to flow in the absence of superconducting phase bias. We relate the associated field-tunable phase shift {$\phi_0$} in the Josephson relation of such a $\phi_0$-junction to the existence of a so-called helical superconductivity, which may result from the interplay of the Zeeman effect and spin-orbit coupling. We analyze the dependence of the magneto-supercurrent on the junction length and discuss its observability in suitably designed hybrid structures subject to an in-plane magnetic field. 
\end{abstract}

\pacs{74.45.+c, 74.78.Na, 71.10.Pm, 85.25.Cp}

\maketitle

\section{Introduction} 
The topological properties of quantum spin-Hall  insulators (QSHI) manifest themselves in current-carrying helical edge states, characterized by a locking of their group velocity to the spin orientation~\cite{Kane_2005,Bernevig_2006,koenig_2006,InAsGaSb-th,InAsGaSb-exp}. The effective one-dimensional (1D) superconductivity  induced in those edge states by conventional superconductors (S) via proximity effect  is predicted to also be topological~\cite{Fu_2009}. This is expected to lead to a fractional Josephson effect in S-QSHI-S junctions, due to topologically protected Andreev bound states {(ABS)}~\cite{Kitaev_2001,Kwon_2003}.  Recent experiments on HgTe/CdTe~\cite{yacoby} or InAs/GaSb~\cite{kouwenhoven} as QSHI, contacted with conventional superconducting leads,  demonstrated that the Josephson current is mainly carried by edge states in the QSH regime.  However, a clear signature of the topological superconductivity induced on edge states is still lacking.  

The role of a Zeeman field transverse to the spin quantization axis of the edge states has been discussed intensively. In the \lq\lq bulk\rq\rq\ it may induce a transition from topological to topologically-trivial superconductivity~\cite{Kitaev_2001,rev-Alicea} whereas a local field in the junction area acts as an effective barrier~\cite{Fu_2009}.  Much less is known about the role of a Zeeman field parallel to the spin quantization axis. An anomalous Josephson effect, that is, a supercurrent flow when no superconducting phase bias is applied to the junction, was calculated in S-QSHI-S junctions with a local Zeeman field applied in the junction area~\cite{linder_2011}. Similar results in junctions with 3D topological insulators were also obtained~\cite{nagaosa_2009,nussbaum}. These systems realize so-called $\phi_0$-junctions, where the current-phase relation has a phase shift $\phi_0$ that is tunable with the external magnetic field. The effect found in those works disappears in the short-junction limit. However, experiments~\cite{yacoby,kouwenhoven} involve a magnetic field applied to the entire system. We show that, in this case, the anomalous Josephson effect depends both on the field in the superconductor  and the field in the junction area via two different though related mechanisms. As a result, we find an anomalous current in junctions of any length, with an amplitude that can be as large as their critical current in the absence of a magnetic field. Specifically, we show that the effect of the field in the superconductor allows one to probe whether superconductivity is indeed induced in the edge states.

Note that an anomalous Josephson effect may occur in any junction with broken time-reversal symmetry~\cite{Geshkenbein}. Indeed, $\phi_0$-junctions were discussed previously in a variety of topologically-trivial systems, mainly in the presence of spin-orbit coupling and a Zeeman field applied to the junction area~\cite{Krive_2004,buzdin_2008,reynoso_2008,nazarov_2013,nazarov_2014,egger_2009,campagnano_2014}, but also in superconductor-ferromagnet hybrid structures~\cite{braude_2007,eschrig_2008}. In the former case, the predicted effect is small. Namely, when both spin-orbit-induced helical bands cross the Fermi level, a partial compensation between the two helicities occurs and only a residual effect proportional to the mismatch in their densities of state remains. In particular, in a 1D system, such a mismatch requires deviations from a parabolic spectrum~\cite{mironov_2014}. The predicted anomalous Josephson effect has not been observed experimentally in these systems, so far. 
The large anomalous current we find in S-QSHI-S junctions is a direct consequence of the helical nature of the edge states. It should be observable in setups similar to those studied in Refs.~\onlinecite{yacoby,kouwenhoven}.

The outline of the article is as follows: We introduce the model in Sec.~\ref{sec:model}.  Then, we study the anomalous Josephson effect in a S-QSHI-S junction along a single edge in Sec.~\ref{sec:CPR}. In particular, we determine the current-phase relation as a function of the external magnetic field and the junction length. We turn to the observability of the effect in S-QSHI-S junctions where both edges contribute in Sec.~\ref{sec:2edges}. We argue that Josephson junctions of unequal lengths should be realized in order to observe the effect. Finally, in Sec.~\ref{sec:discussion}, we discuss that the anisotropy of the gyromagnetic tensor should allow for the observation of the effect with an in-plane magnetic field.   We also show that the effect is robust with respect to a small misalignement between the applied field and the spin quantization axis.

\section{The model}
\label{sec:model}

A Josephson junction formed along one of the edges of a QSHI can be described by the 1D Bogoliubov-de Gennes Hamiltonian~\cite{Fu_2009,rev-Alicea}
\begin{eqnarray}
\label{HBdG}  
H= (v_F p_x \sigma_3 -\mu )\tau_3 -h\sigma_3  +  \Delta(x) \tau_{+} + {\Delta^*(x) \tau_{-}} \ .
\end{eqnarray}
Here, $v_F$ is the Fermi velocity, $x$ and $p_x$ are the position and momentum operators, respectively, $\mu$ is the Fermi energy, and $h>0$ is a Zeeman field along the spin quantization axis. (The effect of a perpendicular component of the field will be discussed below.) The superconducting gap induced by conventional superconducting leads is given as $\Delta(x)=\Delta_0 [e^{-i{\phi}/2} \Theta(x-L/2)+e^{i{\phi}/2}\Theta(-x-L/2)]$, where $\Delta_0$ is the magnitude of the gap, $\phi$ is the phase difference between the two leads, and $L$ is the junction length. Moreover, {$\sigma_i$ and $\tau_i$} are Pauli matrices acting in spin and Nambu spaces, respectively, and $\tau_\pm=(\tau_1\pm i\tau_2)/2$. {Here we use units where $\hbar=k_B=1$}.

The role of the magnetic field within the superconducting regions is most easily understood by considering {first} a  1D \enquote{bulk} superconductor, {i.e., by setting $L=0$ and $\phi=0$ in Eq.~\eqref{HBdG}}. The Zeeman term induces a momentum mismatch $2h/v_F$ between left- and right-moving states at the Fermi level which may be gauged out using the unitary transformation $H\to U^\dagger H U$ with $U=e^{i(h/v_F)x\tau_3}$. However, this gauge transformation modifies the order parameter, $\Delta_0\to\Delta_0e^{-2i(h/v_F)x}$. Thus, for a uniform order parameter, $\Delta_0={\rm const}$, one obtains a current-carrying excited state, whereas the  ground state would require a spatially modulated order parameter with   wavevector $q$, $\Delta_0(x)=\Delta_0e^{iqx}$.  

Indeed, the free energy density of the system can be easily computed and depends on the modulation wavector $q$ through an effective field $h_q=h-v_Fq/2$. For details, see the appendix.
At zero temperature, one finds $F(h,q)=F_0+\Delta_0^2/(2\pi v_F)f(h_q/\Delta_0)$, where $F_0$   is independent of $q$  
and $f(x)=x^2+\Theta(|x|-1)[\arccosh x-|x|\sqrt{x^2-1}]$. The supercurrent is obtained using the thermodynamic relation $I=-2e(\partial F/\partial q)$. One readily shows that  the free energy density is minimized and the current is zero for $h_q=0$, corresponding to a modulation wavevector $q=2h/v_F$. Such a modulated or so-called \lq\lq helical\rq\rq ~superconductivity has been studied in higher dimensions~\cite{Edelstein}.

By contrast, if superconductivity is induced by a conventional  bulk superconductor with constant phase, the induced order parameter inherits the  bulk superconducting phase,  and  a modulation is not possible.   
Then $q=0$,  and the superconductivity induced in the edge states carries a current,
\begin{equation}
\label{I}
I(h)= \frac{e}{\pi} \left[ h -\Theta(h-\Delta_0) \sqrt{h^2-\Delta_0^2} \right]\ . 
\end{equation}
This is precisely the current $I(\phi=0)$ that would flow in a short junction, $L\ll\xi$ with $\xi=v_F/\Delta_0$,  at zero phase difference. Thus, the fact that the proximity-induced superconductivity {forces} the system into an excited state yields an anomalous Josephson effect. The anomalous current increases proportionally to $h$ at $h<\Delta_0$ and then decreases as $ I\simeq e\Delta_0^2/(2\pi h)$ at $h\gg \Delta_0$. 

In the following, we extend the result \eqref{I} to arbitrary  junction lengths and temperatures, and study the current-phase relation. Note that the fact that $\Delta_0$ is an induced gap also implies that there is no self-consistency condition and that fields $h>\Delta_0$ are possible as long as $\Delta_0$ is sufficiently smaller than the intrinsic gap of the superconducting leads.

\section{The current-phase relation}
\label{sec:CPR}

We use the formalism of Refs. \cite{beenakker_1991,brouwer_1997,dolcini-giazotto} to obtain the Josephson current  from the Hamiltonian~\eqref{HBdG}, 
\begin{equation}
\label{eq:IJ}
I_J=-4eT\frac{d}{d\phi}\Re\sum_{\nu=0}^\infty \ln
\left[
1-a^2(\omega_\nu-ih)e^{-2(\omega_\nu-ih)/E_L}e^{i\phi}
\right]
 \ .
\end{equation}
Here $\omega_\nu=(2\nu+1)\pi T$ are Matsubara frequencies at temperature $T$, $a(\omega)=i(\omega-\sqrt{\omega^2+\Delta_0^2})/\Delta_0$,
and $E_L= v_F/L$ is the Thouless energy of the junction. Equation~\eqref{eq:IJ} accounts  for the contributions of both the states in the continuum outside the superconducting gap and the {ABS}, whose subgap energies $E_n$
correspond to the poles of the r.h.s of Eq. \eqref{eq:IJ} after analytic continuation, $\omega_\nu\to -iE+0^+$. In particular, the ABS energies {read} 
\begin{equation}
\label{eq:EA}
2\arccos\left(\frac{E_{n}+h}{\Delta_0}\right)-\frac{2(E_{n}+h)}{E_L}=\phi+2\pi n, \hspace{0.5cm} n\in\mathbb{Z}.
\end{equation}
Equation \eqref{eq:IJ} can be used to numerically compute the current-phase relation at arbitrary junction lengths and temperatures. The results at low temperatures and various fields are shown in Fig.~\ref{Fig-current-phase}. The current-phase relation and the corresponding anomalous Josephson current as a function of the magnetic field are shown for a short junction [$L=0.1\xi$, panels (a)-(b)] and a long junction [$L=10\xi$, panels (c)-(d)], respectively. Below we analyze both short and long junctions further, starting with the limit of zero temperature.

\subsection{{Short {\it vs} long junction}}

\begin{figure}
\centering
\includegraphics[width=\linewidth]{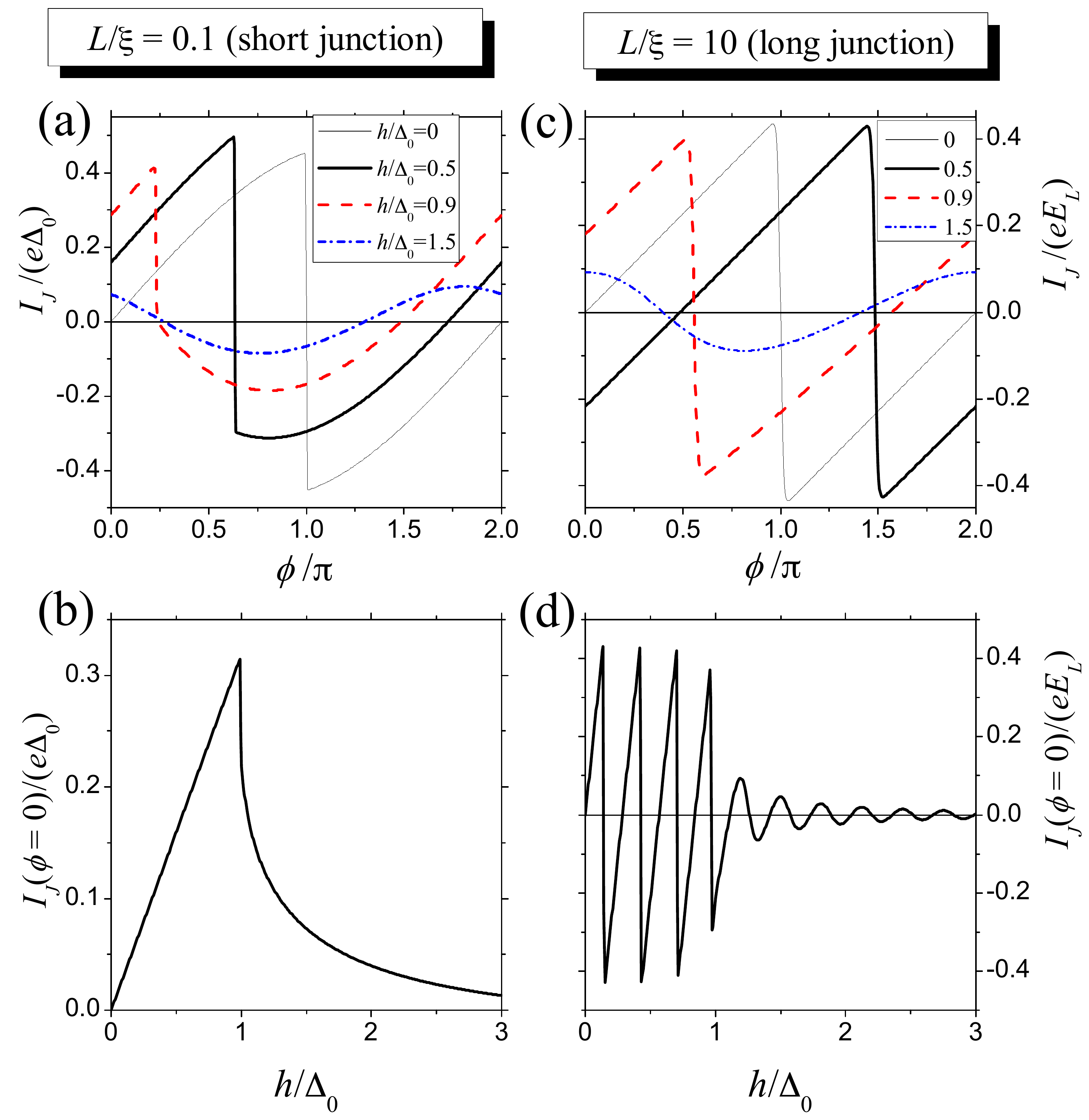}  
\caption{\label{Fig-current-phase} \small Anomalous Josephson effect in short [$L=0.1\xi$, panels (a) and (b)] and long [$L=10\xi$, panels (c) and (d)] S-QSHI-S junctions. Panels (a) and (c) show the current-phase relation  at temperature  $T/\Delta_0=10^{-3}$  for different values of the applied  magnetic field $h$. Panels (b) and (d) show the anomalous Josephson current at $\phi=0$  {as a function of $h$}. } 
\end{figure}

In the short junction limit, $\Delta_0\ll E_L$, we find that the continuum states are essential in determining the current-phase relation (in contrast with conventional {short} junctions, where the supercurrent is carried by ABS only~\cite{beenakker_1991}). Evaluating Eq.~\eqref{eq:IJ} at $\phi=0$, one readily recovers the result \eqref{I} which is a pure continuum contribution. At finite $\phi$, the junction accommodates for a single bound state with energy {$E_A=\Delta_0 \cos(\phi/2)-h$}. The unique zero-energy solution at $\phi^*=2\arccos(h/\Delta_0)$ for $h<\Delta_0$ is a consequence of the topological nature of the junction. It leads to a jump in the current phase relation, cf. Fig.~\ref{Fig-current-phase}(a), which disappears at $h>\Delta_0$, signaling the transition to a topologically trivial state. 
The current-phase relation can be obtained by expanding Eq.~\eqref{eq:IJ} in harmonics and evaluating each term. Summing up the series, one finds, at $T=0$ and $h<\Delta_0$,
\begin{eqnarray}
I_J(\phi,h)=\frac{eh}{\pi}+\frac{e \Delta_0}{2} \sin \frac{\phi}{2} \,\mbox{sign} \left[ \sin\left(\frac{\phi-\phi^*}{2}\right)\right],\, \, \,\,
\end{eqnarray}
where the two terms correspond to the continuum and the ABS contributions, respectively~\cite{large-h}.   The bulk contribution due to the field in the superconductors yields
an asymmetry between the critical currents in opposite
directions, $I_c^+>I_c^-$. Such an anomalous Josephson effect is, thus, a direct probe of the nature of the induced superconductivity underneath the contacts~\cite{asymmetry}.

We now turn to the long junction limit, $\Delta_0\gg E_L$. At $h<\Delta_0$, Eq.~\eqref{eq:EA} yields a large number of ABS with energies $E_{n}=- E_L \left[{\phi}+\phi_h+2\pi(n+1/2)\right]/2$,
where $\phi_h=2h/E_L+2\arcsin(h/\Delta_0)$. The phase shift $\phi_h$ has two  contributions: The first term is proportional to the junction length and can be traced back to the magnetic field in the junction area, it is dominant for $E_L\ll\Delta_0$. The second term stems from the bulk effect discussed above. Correspondingly, the current-phase relation at $T=0$ is   
\begin{equation} 
I_J(\phi,h)=\frac{e E_L} {2\pi} \left[ {\phi}+ {\phi}_h-{2\pi}\,{\rm Int}\left(  \frac{{\phi}+ {\phi}_h}{2\pi}\right)  \right] \ ,  
\label{long-below}
\end{equation}
which extends the result obtained for long S-QSHI-S junctions in the absence of a magnetic field~\cite{beenakker_2013,crepin_2014}.  
The anomalous Josephson current, thus, displays a sawtooth behavior as a function of the applied magnetic field, which is visible in Fig.~\ref{Fig-current-phase}(d). This is reminiscent of the Little-Parks effect~\cite{LittleParks}, though with a paramagnetic rather than orbital origin. As in the short junction limit, the topological nature of the junction manifests itself in a jump in the current-phase relation when the lowest ABS reaches zero energy, at $\phi^*=(\pi-\phi_h ) \, {\rm mod}\, 2\pi$.

At larger fields, $h>\Delta_0$, 
one finds 
\begin{equation}
I_J(\phi,h)=-\frac{e E_L}\pi  \arctan \left[ \frac{\sin({\phi}+\frac{2h}{E_L})}{e^{2\arccosh\frac h{\Delta_0}}-\cos({\phi}+\frac{2h}{E_L})} \right] .  
\label{long-above}
\end{equation}
As expected, there is no more jump in the current-phase relation, and the anomalous Josephson current is suppressed with increasing field.

The behavior of short and long junctions is quite different. In short junctions, the anomalous Josephson effect stems from the magnetic fields in the leads. By contrast, in long junctions, the dominant contribution at small fields comes from the magnetic field in the junction area. In general,  both the field in the leads and in the normal part of the junction contribute. Note, however, that the slope of the anomalous Josephson current as {a} function of the field near $h=0$ is the same in junctions of any length.
 
Namely,
\begin{equation}
\frac{\partial I_J(\phi=0,h)}{\partial h}\Big|_{h=0}=\frac e{\pi}.
\end{equation}
This universal result is a consequence of the helical nature of the QSHI edge states.

\subsection{Finite temperature effects} 

Finite temperatures smear out the sharp features in the current-phase relation. At  $T\gg{\rm min}[\Delta_0,E_L]$,  a sinusoidal behavior,
\begin{equation}
I_J(\phi,h)=I_c(\Delta_0,h,T,E_L)\sin[\phi+\phi_0(h,T,E_L)],
\end{equation}
is found for both short and long junctions. In particular, for short junctions,  we find $I_c=e\Delta_0^2\left|\psi_1(z)\right|/(4\pi^2 T)$ and $\phi_0={\rm arg}\{ \psi_1(z)\}$, where $\psi_1$ is the digamma function and $z=1/2-ih/(2\pi T)$, such that the phase shift  increases from 0 to $\pi/2$ with the field. In long junctions, we find $I_c=4eT\exp[-2\pi T/E_L] |a(2\pi T-ih)|^2$ and $\phi_0=2h/E_L+2\,{\rm arg}\{a(2\pi T-ih)\}+\pi$. For $h\ll T\ll\Delta_0$, one obtains $|a(2\pi T-ih)|=1$ and ${\rm arg}\{a(2\pi T-ih)\}=-\pi/2$, whereas  for $T\ll\Delta_0\ll h$, one obtains
$|a(2\pi T-ih)|=\Delta_0^2/(2h)^2$ and ${\rm arg}\{a(2\pi T-ih)\}=0$.

\section{Double Junctions} 
\label{sec:2edges}

\begin{figure}
\centering
\includegraphics[width=0.85\linewidth]{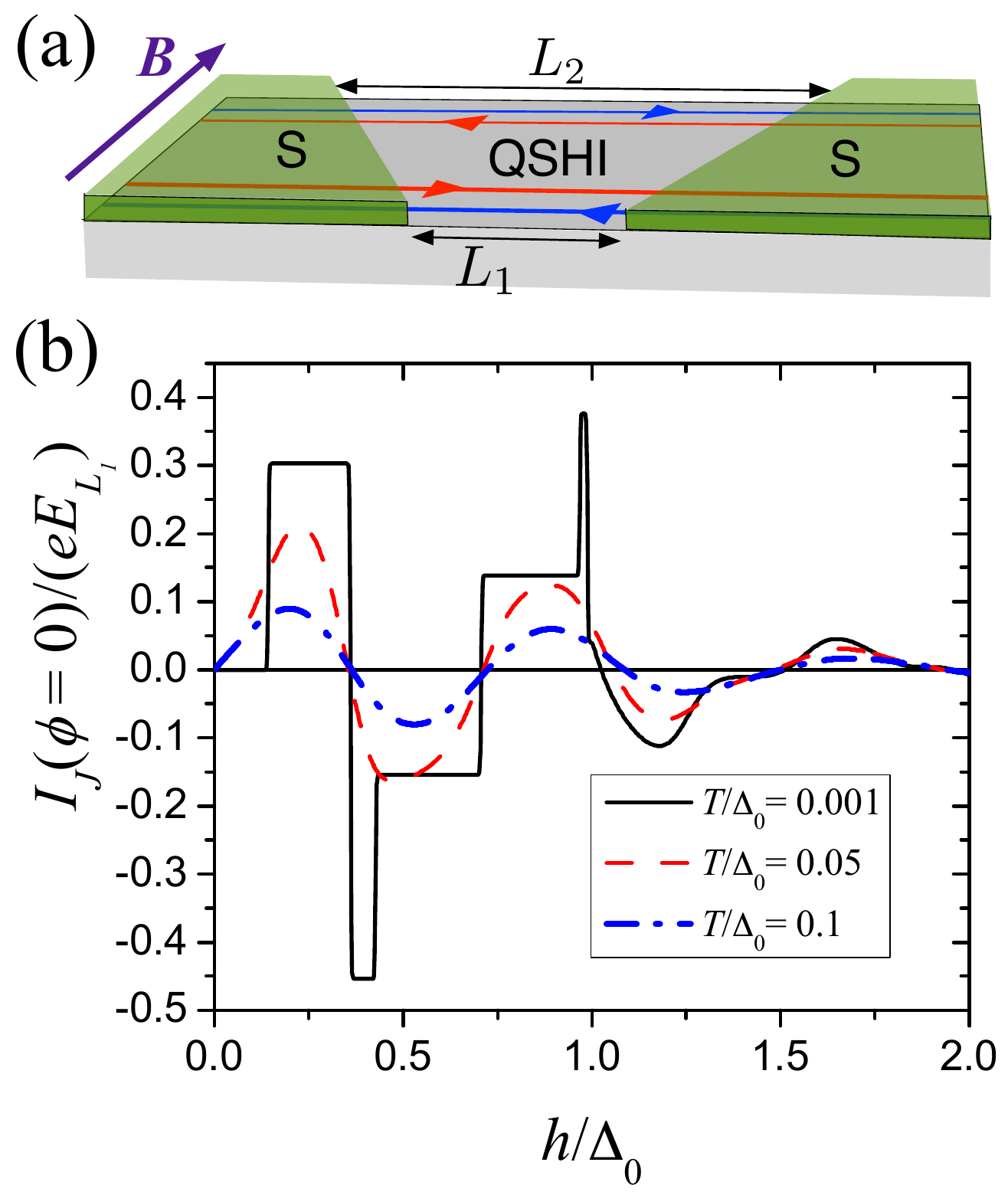}  
\caption{\label{Fig2} \small (a) Proposed setup to detect the $\phi_0$-junction behavior in a S-QSHI-S  hybrid system: The magnetic field ${\bf B}$ is applied in the plane of the quantum well. The edge states on both sides of the sample contribute to the Josephson current. A net anomalous Josephson effect remains, if the junctions have unequal lengths, $L_1\neq L_2$. (b) The anomalous Josephson current at $\phi_0$ {as a function} of the Zeeman splitting $h={\mu_B g_{\rm eff} |{\bf B}|/2}$ in the case $L_2=10\xi$ and $L_1=L_2/3$ for various temperatures.}

\end{figure}

When creating a Josephson junction with a QSHI, typically both edges of the QSHI contribute to the Josephson current~\cite{yacoby,kouwenhoven}. If the  width of the QSHI is sufficiently large, the system may be described as two junctions in parallel and their contributions may be computed separately. Then, as the two edges of the QSHI have opposite helicities, the  contribution of the second edge   can be accounted for by another copy of  Hamiltonian  (\ref{HBdG}) with $h\to-h$. The corresponding current {is} $I_J(\phi,-h)=-I_J(-\phi,h)$.  Similar to the case of conventional $\phi_0$-junctions discussed in the Introduction, adding the current contributions of the two edges leads to a (partial) compensation of their anomalous Josephson currents. 
Here, however, the spatial separation of the two helicities makes an important difference.  
Only if the two junctions on either side of the sample have the same length the compensation is exact, and we obtain the conventional result, $I_J^{\rm total}(\phi=0,h)=0$. However, if the two junctions have unequal lengths, as shown schematically in Fig.~\ref{Fig2}(a), the compensation is only partial and a residual effect remains. This residual effect is a signature that the Josephson current is carried by  {\em helical} edge states. The dependence of the anomalous Josephson current at $\phi=0$ is plotted as a function of $h$ for various temperatures in Fig. \ref{Fig2}(b).  Note that is straightforward to take into account the additional phase shift between the two edges due to the orbital effect of the field, if the junction area encloses a magnetic flux~{\cite{yacoby,kouwenhoven,tkachov}}. 

\section{Discussion and conclusions}  
\label{sec:discussion}

We now turn to the conditions of applicability of our model \eqref{HBdG}. According to the Bernevig-Hughes-Zhang (BHZ)  model for inverted electron-hole bilayers~\cite{Bernevig_2006}, the spin quantization axis points along the growth direction of the heterostructure, and the Zeeman field needed to obtain a $\phi_0$-junction would originate from an out-of-plane magnetic field. This configuration would most likely suppress superconductivity in the leads. However, in real systems, the BHZ model should be supplemented with bulk inversion asymmetry (BIA) and Rashba spin-orbit coupling terms~\cite{review-JPSJ}. Those terms result in a tilt of the quantization axis toward the quantum well plane, thereby allowing  for the generation of the Zeeman field appearing  in Eq.~\eqref{HBdG} with an in-plane magnetic field.

{To obtain} the dependence of this Zeeman field on an external magnetic field $\bf B$, one needs to determine the effective Zeeman term in the Hamiltonian for the helical edge states,  $H_Z=\sum_{i=1,2,3}({\bf t}_i\cdot{\bf B})\sigma_i$. Here ${\bf t}_i$ are vectors that can be computed within the extended BHZ model~\cite{qi_2008}. If the field is applied along the direction ${\bf t}_1\times{\bf t}_2$, {no spin gap opens in the  edge excitation spectrum,} and one obtains Eq.~\eqref{HBdG} with an exchange field $h=-{\bf t}_3\cdot {\bf B}$. Specific values for ${\bf t}_{1,2}$ in a $7 {\rm nm}$-thick HgTe/CdTe quantum well~\cite{qi_2008} indicate that the preferential direction lies close to the plane and perpendicular to the edges. The effective $g$-factor  $g_{\rm eff}$ is expected to be fairly large~\cite{review-JPSJ}. 
\begin{figure}
\centering
\includegraphics[width=0.9\linewidth]{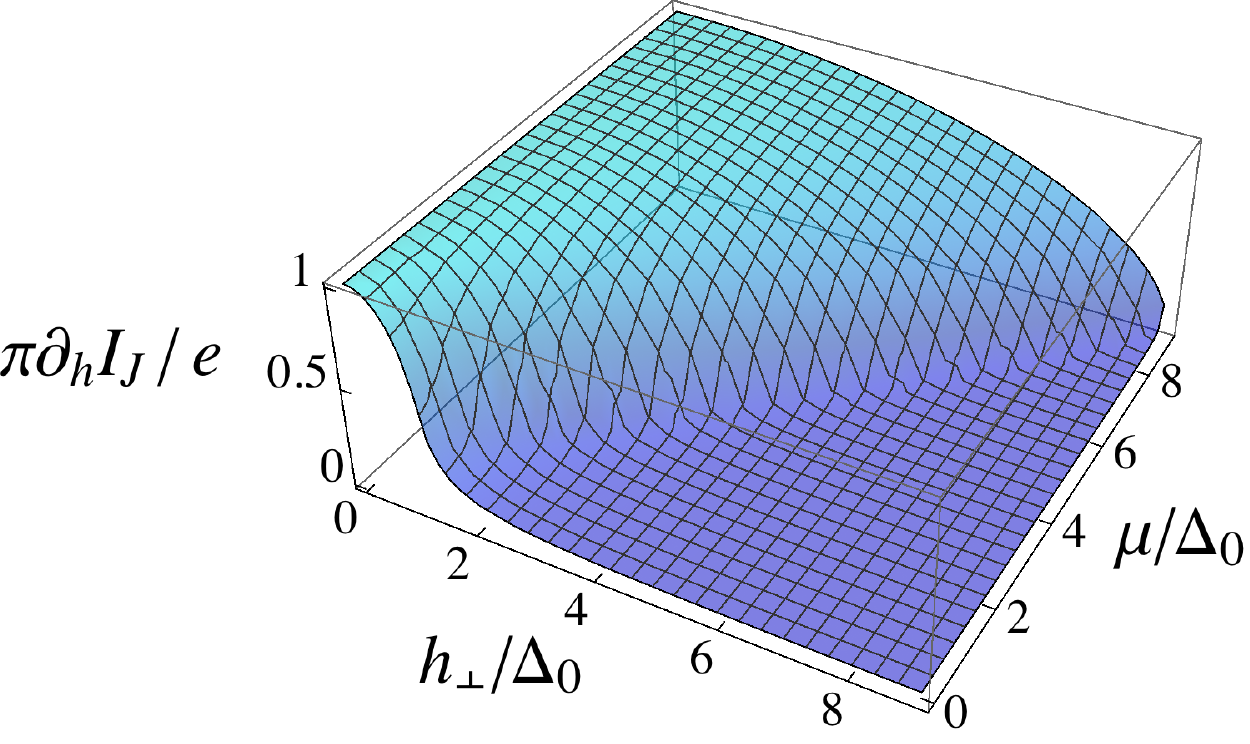}  
\caption{\label{Fig3} \small The slope of the anomalous Josephson current in short junctions as a function of the magnetix field $h_\perp$ perpendicular to the spin quantization axis and the chemical potential.  The slope remains close to its quantized value as long as $h_\perp\ll\sqrt{\Delta^2_0+\mu^2}$.  }
\end{figure}

Experimental studies of the magneto-conductance anisotropy show indeed that the conductance is hardly affected by an in-plane field, while a perpendicular magnetic field leads to a large suppression~\cite{koenig_2006}. These results also show that the topological protection against backscattering, {although in principle} not guaranteed {when} time-reversal symmetry is broken by an in-plane magnetic field, is {in practice} approximately conserved.

Our results are robust with respect to the opening of a small spin gap due to  a misalignment between the field and the spin quantization axis. The anomalous Josephson current in a short junction for an arbitrary direction of the applied magnetic field is derived in the {Appendix}. In Fig.~\ref{Fig3}, we show $\partial_{h} I_J(\phi=0,h)|_{h=0}$ as a functions of the field $h_\perp$ perpendicular to the spin quantization axis and the chemical potential $\mu$. The slope remains close to its quantized value as long as $h_\perp\ll\sqrt{\Delta_0^2+\mu^2}$. By contrast, when $h_\perp>\sqrt{\Delta_0^2+\mu^2}$, the system becomes topologically trivial, and the effect disappears rapidly. 

In conclusion, we have demonstrated that the helical nature of the QSHI edge states leads to an  anomalous Josephson effect in S-QSHI-S junctions subject to a magnetic field. The resulting anomalous supercurrent, flowing at zero phase difference between the two superconducting leads, is field tunable.  Both the field in the superconductor and in the junction area contribute to the effect and probe the helical nature of the edge states in the corresponding parts of the system. We also discussed how to observe this effect using hybrid structures based on available QSHI realizations, analyzing the contributions of both edges, the required magnetic field direction as well as the stability of the effect with respect to a finite chemical potential and a misalignment of the magnetic field and the spin quantization axis. Similarly, we expect a pronounced anomalous Josephson effect in junctions based on nanowires with strong spin-orbit coupling~\cite{lutchyn,oreg}, when they are in the topological regime.

\acknowledgments

FD greatly acknowledges financial support by the Visiting Scientist Program of the Centre de Physique Th\'eorique de Grenoble-Alpes (CPTGA) and the FIRB 2012 project HybridNanoDev (Grant No.~RBFR1236VV).
MH and JSM acknowledge support by ANR through
grants ANR-11-JS04-003-01 and ANR-12-BS04-0016-03,
and an EU-FP7 Marie Curie IRG.

\appendix*

\section{Free energy and anomalous Josephson current of short junctions}
\label{app-free}

In this appendix, we derive an expression for the anomalous current, $I(h)=I_J(\phi=0,h)$, carried by one of the edges of a short S-QSHI-S junction for an arbitrary orientation of the Zeeman field, ${\bf h}=h{\bf e}_3+h_\perp{\bf e}_1$, where ${\bf e}_1$ and ${\bf e}_3$ are perpendicular unit vectors, and the spin quantization of the helical edge state is along ${\bf e}_3$. 


As we argued in the main text, the anomalous current is obtained as $I=-2e(\partial F/\partial q)|_{q=0}$, where $F$ is the free energy {density} of a \lq\lq bulk\rq\rq~1D superconductor with a spatially modulated superconducting order parameter, $\Delta(x)=\Delta_0 e^{iqx}$, where $q$ is the helical modulation wavector. In the following, we assume $q>0$ for definiteness.

As the linear spectrum of the helical edge states is not bounded from below, the free energy will depend on the large momentum cut-off, $k_c$. To properly treat this cut-off, we write the Hamiltonian in the form
\begin{eqnarray}
\label{H}
{\cal H}&=&\sum_{|k|<k_c}\left[ (v_F k-\mu-h)a^\dagger_k a_k-(v_F k+\mu-h)b^\dagger_k b_k\right.\\
&&\left.-h_\perp(a^\dagger_k b_k+b^\dagger_k a_k)\right]
+\!\!\sum_{|k|<k_c-\frac q2}\!\!\Delta_0 a^\dagger_{k+\frac q2} b^\dagger_{-k+\frac q2}+{\rm h.c.}\ .\nonumber
\end{eqnarray}
Here, $a_k$ and $b_k$ are annihilation operators for right-moving (spin up) and left-moving (spin down) electrons with momentum $k$ and Fermi velocity $v_F$ \cite{remark}. Without loss of generality, we choose $\Delta_0$ to be real and positive. As the pairing term couples states with different momenta, the effective cut-off becomes $q$-dependent. Thus, when introducing the Bogoliubov-de Gennes Hamiltonian (cf. Eq.~(1) in the main text), additional cut-off dependent terms have to be kept. Namely, the Hamiltonian \eqref{H} may be split into two parts
\begin{eqnarray}
\label{H2}
{\cal H}=\frac 12 \sum_{|k|<k_c}\Gamma^\dagger_k H_k\Gamma_k+{\cal H}^>,
\end{eqnarray}
where 
\begin{equation}
\label{BdG}
H_k=(v_Fk\sigma_3-\mu)\tau_3+\Delta_0\tau_1-h_\perp\sigma_1-h_q\sigma_3,
\end{equation}
with $h_q=h-v_Fq/2$, is the Bogoliubov-de Gennes Hamiltonian, which is expressed with the help of Pauli matrices $\sigma_i$ and $\tau_i$ acting in spin and Nambu spaces, respectively, and $\Gamma_k=\left(a_{k+\frac q2}, b^\dagger_{-k+\frac q2} , b_{k+\frac q2}  , -a^\dagger_{-k+\frac q2}\right)^T$. Furthermore,
\begin{eqnarray}
{\cal H}^>&=&
-\sum_{\nu=\pm}\nu\sum_{\nu k_c<k<\nu k_c+q/2}
(\begin{array}{cc} a^\dagger_k  & b^\dagger_k\end{array})
H^>
\left(\begin{array}{c} a_k  \\ b_k \end{array}\right)\quad\\
&&-\frac\mu2 \sum_{|k|<k_c}\nonumber
\end{eqnarray}
with
\begin{eqnarray}
H^>=\left(\begin{array}{cc} v_Fk-\mu-h &0 \\0&-v_Fk-\mu+h \end{array}\right)
\end{eqnarray}
is the contribution from energies close to the cut-off, where modifications of the spectrum due to superconductivity as well as the transverse field are negligible. Its contribution to the free energy density is obtained by taking the expectation value of ${\cal H}^>$ in the ground state of the system, yielding
\begin{eqnarray}
F^>&=&\frac1L \sum_{-k_c<k<-k_c+q/2}(v_Fk-\mu-h)\\
&&- \frac1L \sum_{k_c<k<k_c+q/2}(-v_Fk-\mu+h)-\frac{\mu k_c}\pi\nonumber\\
&=&\frac 1{2\pi v_F}\left(h_q^2-h^2\right)-\frac {\mu k_c}\pi.\nonumber
\end{eqnarray}
Let us now turn to the Bogoliubov-de Gennes Hamiltonian $H_k$. It has four eigenenergies, $E_i(k)$ with $i=1,2,3,4$, determined through the equation
\begin{eqnarray}
\label{en}
(E^2-\lambda_{+,k}^2)(E^2-\lambda_{-,k}^2)+8\mu v_F k h_qE=0
\end{eqnarray}
with
\begin{eqnarray}
\label{en-pm}
\lambda_{\pm,k}^2&=&v_F^2k^2+\mu^2+h_q^2+h_\perp^2+\Delta_0^2\\
&&\pm2\sqrt{v_F^2k^2(\mu^2+h_q^2)+(\mu^2+\Delta_0^2)(h_q^2+h_\perp^2)}.\nonumber
\end{eqnarray}
The general expression for the free energy density reads $F=(1/L)\sum_{|k|<k_c}\sum_{i;E_i>0}E_i[f_T(E_i)-1/2]+F^>$, where $f_T(E)$ is the Fermi function at temperature $T$. Note that the solutions obey the relation $E_i(k)=-E_i(-k)$. Thus, at zero temperature, the free energy density of the system takes the form
\begin{eqnarray}
\label{F}
F=-\frac 1{4\pi} \int\limits_0^{k_c} dk\; \sum_i |E_i(k)| + F^>.
\end{eqnarray}
The current is then given as
\begin{eqnarray}
\label{res}
I(h)=-\frac {ev_F}{4\pi} \int\limits_0^{k_c} dk\; \sum_i \partial_h |\bar E_i(k)|+ \frac e\pi h,
\end{eqnarray}
where $\bar E_i(k)$ are the eigenenergies of the system at $q=0$.
Analyzing Eqs.~\eqref{en} and\eqref{en-pm}, we note that zero-energy solutions exist if $h^2+h_\perp^2=\Delta_0^2+\mu^2$ or $|h|>\Delta_0$. The topologically non-trivial region corresponds to low fields, {$h^2+h_\perp^2<\Delta_0^2+\mu^2$} and $|h|<\Delta_0$.

We further analyze the result \eqref{res} in two limiting cases, namely at $h_\perp=0$ and for arbitrary $h_\perp$ in the limit $h\to0$.

\subsection{Supercurrent for a Zeeman field parallel to the spin quantization axis}

When the Zeeman field is parallel to the quantization axis, $h_\perp=0$, the eigenenergies are
\begin{equation}
E_{s_1s_2}(k)=s_2\sqrt{(v_Fk+s_1 \mu)^2+\Delta_0^2}-s_1h_q\ ,
\end{equation}
where $s_1,s_2=\pm1$. Evaluation of Eq.~\eqref{F} at zero temperature then yields\begin{eqnarray}
F&=&F_0+\frac1{2\pi v_F}\Bigg\{
h_q^2\\
&&+\theta(|h_q|-\Delta_0)\left[\Delta_0^2 \arccosh\frac{h_q}{\Delta_0}
-|h_q|\sqrt{h_q^2-\Delta_0^2}\right]
\Bigg\}\ ,\nonumber
\end{eqnarray}
with 
\begin{eqnarray}
F_0&=&
-\frac1{2\pi v_F}\
\Bigg\{
(v_Fk_c+\mu)^2\\
&&+\Delta_0^2\left[\frac 12+\ln\left(\frac{2v_Fk_c}{\Delta_0}\right)\right]+h^2
\Bigg\}\nonumber
\end{eqnarray}
that does not depend on $q$, in agreement with the expression given in the main text. Evaluating the anomalous current $I(h)$ using Eq.~\eqref{res}, we then obtain Eq.~(2) in the main text. In the topological regime, $h<\Delta_0$, the large anomalous current is given as $I(h)=(e/\pi)h$.

\bigskip

\subsection{Supercurrent response to a small Zeeman field along the spin quantization axis}

When the field component along the spin quantization is small, $h\to0$, we may evaluate the eigenenergies perturbatively in $h_q$. At $h_q=0$, we obtain $E_{s_1s_2}^{(0)}(k)=s_2E_{s_1}(k) $ with
\begin{eqnarray}
E_{s_1}(k)&=&\Bigg(v_F^2k^2+\mu^2+\Delta_0^2+h_\perp^2\\
&&+2s_1\sqrt{v_F^2k^2\mu^2+(\mu^2+\Delta_0^2)h_\perp^2}\Bigg)^{1/2}\ .\nonumber
\end{eqnarray}
Making use of perturbation theory to obtain the  correction $\delta E_{s_1s_2}(k)$ to the energy $E_{s_1s_2}^{(0)}(k)$, we obtain the free energy density up to quadratic order in $h_q$,
\begin{eqnarray}
F&=&F_1+\frac{h_q^2}{2\pi v_F}\Bigg\{1\\
&&-v_F h_\perp^2\int\limits_{-k_c}^{k_c} dk\,\frac{[E_{+}(k)+E_-(k)]^2-4(\mu^2+\Delta_0^2)}{E_{+}(k)E_-(k)[E_{+}(k)+E_-(k)]^3}\Bigg\}
\ ,\nonumber
\end{eqnarray}
where $F_1$ does not depend on $q$. As the integral converges at large momenta, we may take the limit $k_c\to\infty$.
The slope of the anomalous current in the limit $h \to 0$ is, thus,
\begin{eqnarray}
\label{slope}
\left.\frac{\partial I}{\partial h}\right|_{h=0}&=&
\frac e\pi\Bigg\{1\\
&&\!\!\!\!\!\!\!\!\!\!\!\!-v_Fh_\perp^2\int\limits_{-\infty}^\infty dk\,\frac{[E_{+}(k)+E_-(k)]^2-4(\mu^2+\Delta_0^2)}{E_{+}(k)E_-(k)[E_{+}(k)+E_-(k)]^3}
\Bigg\}
\ .\nonumber
\end{eqnarray}
The dependence of the slope \eqref{slope} on the chemical potential $\mu$ and the transverse field $h_\perp$ is shown in Fig.~3 of the main text. A large signal,  $(\partial I/\partial h)|_{h=0}\approx e/\pi$, is obtained deep in the topological region, when $h_\perp^2\ll \Delta_0^2+\mu^2$.



\begin{thebibliography}{99}
\bibitem{large-h}
For completeness, we provide the expression for the current-phase relation at $T=0$ and $h>\Delta_0$:
\begin{eqnarray*}
\qquad I_J(\phi,h)&=&\frac e\pi \left[h-\sqrt{h^2-\Delta_0^2}\right.
\\
&&\left.- \Delta_0 \sin \frac\phi 2{\rm arctan}\left(\frac{\Delta_0} {\sqrt{h^2-\Delta_0^2}}\sin \frac\phi 2\right)\right].
\end{eqnarray*}
At $h\gg\Delta_0$, it simplifies to  $I(\phi)\approx e\Delta_0^2/(2\pi h) \cos \phi.$
\bibitem{remark}
The model {neglects} a possible rotation of the spin quantization axis with momentum $k$, assuming that it takes place on much a larger energy scale than the superconducting gap.
\end{thebibliography}
\end{document}